\documentstyle[12pt]{article}

\begin{document}

\title{Relative phase change during quantum operation \footnote{Invited 
talk presented at First Feynman Festival on Quantum Computation
at University of Maryland, College Park from August 23-28th, 2002 under the
title ``Quantum phase during quantum operation''.} }

\author{Arun K. Pati$^{(*)}$ \\
Institute of Physics, Sainik School Post,\\
Bhubaneswar-751005, Orissa, India\\
$^{(*)}$ School of Informatics, University of Wales,\\ 
Bangor LL 57 1UT, UK}

\date{\today}

\maketitle
\def\ra{\rangle}
\def\la{\langle}
\def\ver{\arrowvert}

\begin{abstract} 
Quantum operations represented by completely
positive maps encompass many of the physical processes 
and have been very powerful in describing quantum computation and
information processing tasks. We introduce the notion of relative 
phase change for a quantum system undergoing quantum operation. We find
that the relative phase shift of a system not only depends on the
state of the system, but also depends on the initial state of the
ancilla with which it might have interacted in the past.
The relative phase change during a sequence of quantum operations is shown
to be non-additive in nature. This property can attribute a `memory' to a 
quantum channel. Also the notion of relative phase shift helps 
us to define what we call `in-phase quantum channels'.
We will present the relative phase shift for a qubit undergoing 
depolarizing channel and complete randomization and discuss their
 implications.
\end{abstract} 

\maketitle

\newcommand{\nc}{\newcommand}
\nc{\rnc}{\renewcommand}
\nc{\ket}[1]{| #1 \rangle}
\nc{\bra}[1]{\langle #1 |}
\nc{\proj}[1]{\ket{#1}\bra{#1}}
\nc{\braket}[2]{\langle #1| #2 \rangle}
\nc{\hilb}{\mathcal{H}}
\nc{\inprod}[2]{\braket{#1}{#2}}
\def\id{{\mathbf 1}}
\rnc{\vec}[1]{\boldsymbol{#1}}

\vskip 2cm

\newpage

\section{Introduction}
In quantum theory a physical system $S$ is associated with a complex, 
separable Hilbert space ${\cal H}^D$, $D$ being the dimension of the Hilbert 
space which can be finite or infinite. An isolated system is described by a 
pure state which is a normalized vector $| \psi \ra \in {\cal H} $ with
$\la \psi| \psi \ra=1$. Two states are equivalent, i.e., $| \psi \ra \sim
| \psi '\ra$ if $| \psi' \ra = e^{i\alpha}| \psi \ra$ with $\alpha$ real. They
represent the same physical state of a system. Transformations that are 
{\em norm preserving, one-to-one and onto} are unitary, i.e., if
$U: | \psi \ra \rightarrow | \phi \ra$ satisfying above conditions for all
$| \psi \ra, | \phi \ra \in {\cal H}$ then
$| \phi \ra=U | \psi\ra$ with $UU^{\dagger}=U^{\dagger}U=I$. The amplitude for
such a transformation is given by the {\em inner product} function
$\la \psi| \phi \ra =\la \psi|U| \psi \ra$ which is a complex number having
a modulus and a phase. In the quantum world this phase of the amplitude 
of a process is what it makes so different from that of a classical process. 
Though absolute phase has no physical significance, the relative phase of a 
quantum state with respect to another is an important object of study. 
In fact, the relative phase is the most crucial quantity in quantum 
interference which according to Feynman contains the only mysteries 
\cite{feynman}. If we have a quantum system $S$
described by a pure state $| \psi(0)\ra$ at initial time and the state of the
system is $|\psi(t)\ra$ at a later time, then the relative phase shift $\Phi$
during such an evolution is given by the argument of the inner product
function, i.e.,

\begin{equation}
\Phi= {\rm Arg} \langle \psi(0)| \psi(t)\rangle
\end{equation}
Surprisingly, this simple definition of comparing phase of two
distinct states was suggested by Pancharatnam only in 1956 in studying 
interference of polarization of light rays \cite{panch}. He has 
introduced the notion of two states being {\em in-phase}. Accordingly, 
two non-orthogonal states are said to be in phase if the inner-product 
function $\langle \psi(0)| \psi(t)\rangle$ is real and positive.

In recent years we have learned that the the relative phase shift of a 
quantum system can be of various origin, namely, it can be purely dynamical
or it can be purely geometric or both. The discovery of pure geometric
phase shift in the context of light goes back to Pancharatnam. In quantum 
mechanical context it was discovered by Berry in cyclic
adiabatic evolutions \cite{berry84}. Presently there is an immense 
interest in holonomy effects in quantum mechanics and has led to many 
generalizations of the notion of geometric phases 
\cite{aharonov87,samuel88,aitchison92,ms,akp,akp1}.

When a pure state evolves in time (the evolution need not be adiabatic, 
cyclic, unitary, and even Schr{\"o}dinger type) then it traces an open curve 
in the Hilbert space, i.e., $\Gamma : t \in [0,t]$ with $\ver \psi \ra 
\rightarrow \ver \psi(t) \ra$. If we project the evolution curve by a 
projection map $\Pi: \Gamma \rightarrow {\bar \Gamma}$, we have an open 
curve in the projective Hilbert space ${\cal P}$. During the 
evolution it acquires a relative phase which can be called the total phase. 
The total phase can always be 
written as sum of a dynamical phase  

\begin{equation}
\Phi_D= -i \int ~~dt  \la \frac{\psi(t)}{||\psi(t)||} \ver 
\frac{d}{dt} \large (\frac{\psi(t)}{||\psi(t)||} \large) \ra
\end{equation}
and a geometric phase given by
\begin{equation}
\Phi_G= i \int ~~ \la \frac{\chi(t)}{||\chi(t)||} \ver 
{\Large d} (\frac{\chi(t)}{||\chi(t)||} \large) \ra,
\end{equation}
where $\ver \chi(t) \ra$ is a ``reference-section'' introduced in 
\cite{akp,akp1}.  In fiber bundle formulation the inner product of 
``reference-section'' with its path derivative gives the {\em connection 
one-form} whose line integral provides the generalized geometric phase 
\cite{akp}. It can be shown that the above geometric phase is local gauage 
invariant, independent of the detailed dynamics, independent of 
re-parametrization, but depends only on the geometry of the evolution 
path ${\bar \Gamma}$ in the projective 
Hilbert space of the quantum system. The above geometric phase can reduce 
to Berry phase \cite{berry84}, Anandan-Aharonov phase \cite{aharonov87}, 
Samuel-Bhandari phase \cite{samuel88}, Aitchison-Wanelik phase 
\cite{aitchison92}, Mukunda-Simon phase \cite{ms} 
under appropriate limits. In the above generalisation, only the 
existence of an inner product and continuous nature of time evolution, 
all that is used.

One can also introduce a notion of holonomy to mixed 
quantal states \cite{uhlmann86} purely on mathematical ground as well as on
physical ground such as in the context of interferometry \cite{sjoqvist00}.
Usually one may tend to think that a mixture is an incoherent superposition 
of quantum states, so there is no notion of `relative phase'. However, 
it is legitimate to ask what is the 
relative `phase difference' between an initial and final mixtures. 
Unlike in the case of pure states, there are two types of `phases' for 
mixed states and both of them go over to the same expression for pure state 
case \cite{eric}. 
Further differences between Uhlmann phase and Sj{\"o}qvist et al 
phase has been pointed out recently \cite{slater,slater1}. 
The geometric phase may be useful in the context of quantum 
computing as a tool to achieve fault-tolerance 
\cite{jones00,ekert,falci,nazir}. It may be mentioned that for practical 
implementations of geometric quantum computing, it is important 
to understand the behavior of the geometric phase in the presence of 
decoherence. Possibility of measuring geometric phase in
ion traps has also been suggested \cite{gbv}. Since ion traps have been 
proposed as suitable devices for quantum information processing tasks, the 
above proposal may be useful indeed. In another context, a proposal has been
made to produce and observe geometric phase shift for a three-level system 
(a qutrit) in interferometry \cite{barry}.

In this paper we will mention the generalization of relative phase shift
for mixed states undergoing unitary evolution. Using Kraus representation
theorem we will define the relative phase shift for a system (either pure
or mixed) undergoing a quantum operation, namely, a completely positive (CP) 
map. We will also prove that the relative phase shift is non-additive in 
nature, thus attributing a `memory' to a quantum channel. Further, using 
these notion, one can define what I call ``in-phase quantum channel''. 
Through such a channel a quantum state remains in phase with the initial
state. In addition, we will discuss few examples to illustrate these ideas.

\section{Phase shift for mixed states}

In actual physical situation, the state of a quantum system
may not be a pure. This could happen due to variety of reasons. One common
cause is unavoidable interaction with the surrounding that leads to
loss of coherence and thus the state may become a mixture. Other reason
could be that the system of our interest may be a part of a larger
system with composite being in an entangled state. That is to say that 
we do not have access to whole system, but a part of it, thus giving rise 
to a mixed state.

If we have a mixed state $\rho=\sum_{k=1}^{N}w_{k}|k\rangle\langle k|$
that undergoes a unitary time evolution, then
it traces a path $\Gamma: t \in [0,t]$ with $\rho  \rightarrow \rho(t)=
U(t) \rho U^{\dagger}(t)$. For such a situation a computable definition 
of relative phase shift was not known. Sjoqvist {\it et al} 
\cite{sjoqvist00} have generalized the notion of relative phase shift for 
a mixed state undergoing a unitary evolution as follows
\begin{equation}
\Phi= {\rm Arg} {\rm Tr}[\rho U(t)]= \sum_k w_k \la k|U(t)|k\ra.
\end{equation}
This is the relative phase shift between $\rho$ and $\rho(t)$.
That this is a correct prescription was verified using interferometric
techniques. By sending a mixed state  as an input through a Mach-Zehnder 
interferometer with a variable relative $U(1)$ phase $\chi$ in one arm of 
the interferometer (the reference beam) and the other arm (the target beam) 
is exposed to the unitary operator $U(t)$, it was found that the output of 
interference pattern oscillates according to 

\begin{eqnarray} 
I \propto 1 + |{\rm Tr}(U(t) \rho)|\cos[\chi - \arg {\rm Tr}(U(t) \rho)].  
\end{eqnarray} 
Here, one may notice that the interference fringe
produced by varying the phase $\chi$ is shifted by  
$\Phi = {\rm Arg}{\rm Tr}(U(t) \rho)$ and that this shift reduces  
to Pancharatnam's original prescription for pure states. For example, when
$\rho = \ver \psi \ra \la \psi\ver$, we have $\Phi = {\rm Arg}
{\rm Tr}(U(t) \rho)= {\rm Arg}{\rm Tr} \la \psi(0) \ver \psi(t) \ra$.
These two facts are the central properties for $\Phi$ being a  
natural generalization of Pancharatnam's relative phase to  
mixed states undergoing unitary evolution. However, the definition (4) 
does not hold along an evolution path when ${\rm Tr}(U(t) \rho)=0$. 
This situation is similar to the pure state case, where 
Pancharatnam's definition of phase difference also breaks down for two 
orthogonal states \cite{rb,anandan}.

Using this definition it was possible to introduce the notion of
parallel transport condition, holonomy transformation
and connection-form for mixed states undergoing a unitary evolution.
The geometric phase for a mixed state undergoing parallel transportation 
can be expressed as an average connection-form
\begin{equation}
\Phi_G[\Gamma]=  \int  \sum_k w_k i \la \chi_k \ver d \chi_k \ra,
\end{equation}
where $\ver \chi_k\ra$ is the ``reference-section'' for the $k$th 
pure state component in the ensemble. For unitary time evolution of 
pure quantum state the above expression reduces to that of (3) obtained in
\cite{akp} by the present author.

\section{ Phase shift during quantum operation}
 
An open quantum system described by a mixed state may undergo a non-unitary 
evolution. Here, we generalize the notion of relative phase shift
and geometric phase shift to such scenarios. If the system undergoes a 
general quantum operation described by a completely positive map, then what 
would be the relative phase shift? Recently we have made some progress in 
understanding the phase change under a CP map \cite{marie}. 
Geometric phase for non-unitary evolution with continuous 
version of CP map has been independently addressed in \cite{fpn}.
Quantum operation is a very powerful technique in describing many quantum 
information processing tasks. Most of the physical operations such as 
attaching an ancilla, tracing out a subsystem, unitary evolution, 
measurements (such as von Neumann and POVMs) and non-unitary stochastic 
operations may be modeled by quantum operations \cite{nc}. In fact, one can 
imagine that the mixed states result due to some noise acting on the 
system which is also represented by a CP map.

Let ${\cal E}$ be the quantum operation that maps $\rho \rightarrow 
{\cal E}(\rho)= \sum_{\mu} E_{\mu} \rho E_{\mu}^{\dagger}$, where
$E_{\mu}$ is the set of Kraus operators satisfying  the completeness condition 
$\sum_{\mu} E_{\mu}^{\dagger} E_{\mu}=I$. The question we would like to 
answer is {\em what is the relative phase shift between $\rho$ and 
${\cal E}(\rho)$ ?}
Since every CP map has a unitary representation in an extended Hilbert
space, let us imagine that there is an ancilla with an initial state
$|0\ra\la 0|$. The combined state of the system and ancilla is $\rho_s \otimes
|0\ra_a\la 0|$. The combined system undergoes a unitary evolution

\begin{equation}
\rho_s \otimes |0\ra_a\la 0| \rightarrow U(\rho_s \otimes
|0\ra_a\la 0|) U^{\dagger}.
\end{equation} 
Then the evolution of the system is obtained by tracing over this   
ancilla yielding    
\begin{equation} 
{\cal E}(\rho_s)= {\rm Tr}_{a}[ U(\rho_s \otimes
|0\ra_a\la 0|) U^{\dagger}] =\sum_{\mu} 
E_{\mu}\rho_s E_{\mu}^{\dagger}, 
\label{eq:kraus}
\end{equation} 
where the Kraus operators are $E_{\mu} = {_\langle} \mu  |U |0 \rangle_a $
in terms of an orthonormal basis $\{|\mu \rangle\}$, $\mu =
0, \ldots, K-1 \geq N$, of the $K-$dimensional Hilbert space of the
ancilla \cite{kraus83}. In fact, it is sufficient that $K=N^2$. The 
operator elements appearing in (8) are not unique, because different set 
of Kraus operators may give rise to same quantum operation. This is a 
completely positive map 
as it takes density operators into density operators, and also all 
trivial extensions. Conversely, any CP map has a Kraus representation 
of the form Eq.~(\ref{eq:kraus}) \cite{nc,kraus83,preskill}.
 
The relative phase shift of the system can be thought of as the relative phase
shift of the combined system under unitary operation in an enlarged system.
Therefore, tracing out the ancilla gives us 

\begin{eqnarray} 
\Phi & = & {\rm Arg}{\rm Tr}_{s+a} \big[ U \rho_s \otimes |0\ra_a\la 0|) 
\big] \nonumber\\
 & = & 
{\rm Arg} {\rm Tr}_{s} \big[\sum_{\mu} \la \mu |U|0 \ra_a \la 0|\mu\ra 
\rho_s \big] = 
{\rm Arg} {\rm Tr}_{s} \big[ E_0 \rho_s \big]
\label{eq:cpint} 
\end{eqnarray} 
where we have used the orthogonality $\langle 0  | \mu 
\rangle = \delta_{0\mu}$. The quantity $\Phi$ is a natural  
definition of relative phase as it shifts the maximum of the  
interference pattern and reduces to the phase defined in  
\cite{sjoqvist00} for unitarily evolving mixed states.  

Since we have ignored the ancilla, we are legitimate to think that the above
phase is the quantum phase shift for the system undergoing a quantum 
operation. However, the above expression holds true if the initial state of
the ancilla is in one of the orthogonal basis state (say) in the state $|0\ra$.
In next section, we will obtain a general expression for the relative phase
shift when the initial state of the ancilla is in an arbitrary state.

Since phase information has leaked from the system part, the interference
information contained in Eq. (\ref{eq:cpint}) is only partial. The remaining
part may be uncovered by flipping the state of the environment associated with
the reference beam to an orthogonal state $|\mu \neq 0  \rangle$.
This assumes that we have full control over the
ancilla and may be thought of as an extra degree of internal freedom.
This transformation may be represented by the operator
\begin{eqnarray} 
U=\left( \begin{array}{rr} 0 & 0 \\ 0 & 1  
\end{array} \right)\otimes U_{s+a} + 
\left( \begin{array}{rr} e^{i\chi} & 0 \\ 0 & 0  
\end{array} \right)\otimes 1_{s}\otimes  
F_{0 \rightarrow \mu} , 
\end{eqnarray} 
where the first matrix in each term represents the spatial part and
the operator $F_{0 \rightarrow \mu}$ flips
$|0 \rangle$ to $|\mu \neq 0 \rangle$.
The phase shift is determined
by
\begin{eqnarray} 
\Phi_{\mu} & = &  
{\rm Tr}_{s+a}\big[U_{s+a} \rho_s|0   
\rangle \langle 0 |  
F_{0 \rightarrow \mu}^{\dagger} \big]  
\nonumber \\ 
 & = & {\rm Tr}_{s+a}\big[ U_{s+a} \rho_s 
|0  \rangle \langle \mu | \big] =  
{\rm Tr}_{s} \big[ E_{\mu} \rho_s \big]  
\label{eq:cpint2} 
\end{eqnarray}  
for each $\mu = 1, \ldots, K-1$. The phase $\Phi$ and $\Phi_{\mu}$ contain
maximal information about the interference
effect during a quantum operation. For
unitarily evolving mixed states one obtains $\nu_{\mu} = 
\delta_{0 \mu}$, due to orthogonality of the ancilla states, and the 
surviving interference pattern Eq.~(\ref{eq:cpint}) reduces to that of 
\cite{sjoqvist00}.

The above results can be derived by considering purifications. 
We may lift $\rho_s$ to a purified 
state $|\Psi \rangle_{sas'}$ by attaching an ancilla according to
\begin{equation}
|\Psi \rangle_{sas'}=
\sum_k \sqrt{w_{k}} |k \rangle_s |0 \rangle_a |k' \rangle_{s'} 
\end{equation} 
with $\{ |k \rangle_{s'} \}$ a basis in an auxiliary Hilbert space 
of dimension at least as large as that of the internal Hilbert 
space. This state is mapped by the operators $U = U_{s+a}  
\otimes I_{s'}$ and $F = I_{s} \otimes F_{0 \rightarrow \mu} 
\otimes I_{s'}$ in the target and reference beam, respectively, i.e.  
\begin{eqnarray}
|\Psi_{\rm tar} \rangle_{sas'} & = & 
\sum_k \sqrt{w_{k}} \big[ U_{s+a} |k \rangle_s \otimes
|0 \rangle_a \big] \otimes |k' \rangle_{s'} , 
\nonumber \\ 
|\Psi_{\rm ref} \rangle_{sas'} & = & 
\sum_k \sqrt{w_{k}} |k \rangle_s \otimes \big[  F_{0 \rightarrow \mu} 
|0 \rangle_a \big] \otimes |k \rangle_{s'} , 
\end{eqnarray}
The Pancharatnam phase difference between these purified states is
given by
\begin{eqnarray}
\Phi = {\rm Arg} \langle \Psi_{\rm ref} |\Psi_{\rm tar} \rangle = 
{\rm Arg} \sum_k w_{k} \langle k | \langle 0 | 
F_{0 \rightarrow \mu }^{\dagger} U_{s+a} |0 \rangle | k \rangle   
\nonumber \\
={\rm  Arg} \sum_k w_{k} \langle k | E_{\mu} |k \rangle =
{\rm Arg} {\rm Tr}_{s} ( E_{\mu} \rho_s ) 
\label{eq:purification}
\end{eqnarray}
in agreement with Eqs. (\ref{eq:cpint}) and (\ref{eq:cpint2}). 
Hence, we can understand the relative phase shift of a mixed state under 
a completely positive map as a relative phase shift of the purified state 
in a larger Hilbert space under a unitary evolution.

\section{Phase shift and ancilla state}

Here, we will discuss how the relative phase shift for a system 
depends on the choice of the initial state of the ancilla. Suppose that
instead of the initial state of the ancilla being $|0\ra_a\la0|$ it 
is an arbitrary state 
$|A\ra= \sum_{\mu} a_{\mu} |\mu \ra$. Then the general expression for
the phase shift will be 
\begin{eqnarray} 
\Phi & = & {\rm Arg}{\rm Tr}_{s+a} \big[ U \rho_s \otimes |A\ra_a\la A|) 
\big] \nonumber\\
 & = & 
{\rm Arg} {\rm Tr}_{s} \big[\sum_{\mu} \la \mu |U|A \ra \la A|\mu\ra 
\rho_s \big]
\end{eqnarray} 
Notice that this choice of ancilla state gives rise to another
set of Kraus operator and hence a different quantum operation, in general.
The Kraus operators are given by $F_{\mu}=\la \mu |U|A \ra$ and the
quantum operation is described by 
$\rho \rightarrow 
{\cal F}(\rho)= \sum_{\mu} F_{\mu} \rho F_{\mu}^{\dagger}$.
The relative phase shift in this case is given by
\begin{eqnarray} 
\Phi =  {\rm Arg} \sum_{\mu} a_{\mu}^* 
{\rm Tr}_{s} \big[F_{\mu} \rho_s \big].
\end{eqnarray} 
In general (9) and (16) are different. In fact, there is no priori reason
to believe that they are same. This shows that the relative phase shift 
depends on the choice of the initial state of the ancilla.

Alternately, one can express (17) as
\begin{eqnarray} 
\Phi =  {\rm Arg}  
{\rm Tr}_{s} \big[ \rho_s \la A|U|A\ra \big]= 
{\rm Arg} \sum_k w_k \la k |{\cal N}_s |k \ra,
\end{eqnarray} 
where ${\cal N}_s= \la A|U_{s+a}|A\ra$ is a non-unitary operator acting only 
on ${\cal H}_s$. It is intriguing that even if the ancilla may be far apart, 
the quantum operation on the system causing the relative phase shift depends
on what would have been the initial state of the ancilla. On the other hand,
it is not surprising because after all the quantum operation
depend on the choice of the initial state of the ancilla.

One can think that choice of initial state of ancilla is equivalent to
choosing a local unitary transformation on the ancilla, i.e., if $\ver A\ra=
u\ver 0\ra=\sum_{\mu} \la \mu |u|0 \ra |\mu\ra$, then instead of the 
unitary evolution $U$ of combined system, we
have a unitary evolution $U (I \otimes u)$ acting on the system and ancilla. 
In general, they give rise
to two different quantum operations. However, when $U$ and $(I \otimes u)$
commutes, i.e., $U (I \otimes u)= (I \otimes u)U$ then only we have
the same quantum operations. From the Theorem \cite{nc} of Unitary freedom 
of operator-sum representation, we know that two quantum operations are 
equal, i.e., ${\cal E}={\cal F}$ iff 
${\cal F}= \sum_j u_{ij} {\cal E}$ with $u_{ij}$ as complex numbers.
In general $U (I \otimes u) \not= (I \otimes u)U$ and hence,
we have different relative phases. This can provide
a physical means to distinguish two quantum operations.

\section{Non-additive nature of phase}

Suppose we have a quantum system that undergoes a sequence of quantum 
operations described by CP maps. Let us consider two CP maps
${\cal E}$ followed by ${\cal F}$. Thus a 
density matrix $\rho \rightarrow {\cal E}(\rho)=\rho'
\rightarrow {\cal F}(\rho')= \rho''$. 
Also imagine that there is a CP map that can transform $\rho$ directly 
into $\rho''$ via $\rho\rightarrow {\cal G}(\rho)=\rho''$.
Let $\{E_{\mu}\}$, $\{F_{\mu}\}$, and $\{G_{\mu}\}$ are the Kraus elements 
corresponding to quantum operations ${\cal E}$, ${\cal F}$ and ${\cal G}$,
respectively. Let $\Phi_{12}$ be relative phase change
between $\rho$ and $\rho'$, $\Phi_{23}$ be the relative phase 
change between  $\rho'$ and $\rho''$ and $\phi_{13}$ be the relative phase
change between $\rho$ and $\rho''$. Now it can be shown that the phase
shift between $\rho$ and $\rho''$ {\em is not sum} of the phase shifts between 
$\rho$ and $\rho'$,  and $\rho'$ and $\rho''$, i.e., $\Phi_{13} \not=
\Phi_{12}+ \Phi_{23}$. 

One can prove this directly using the definition of relative phases of 
mixed states
with CP maps. But it is simple and illustrative to imagine the unitary 
extension of these sequence of CP maps along with purification of the states
of the system that is of interest to us. Let $U$ be the unitary representation
for the CP map ${\cal E}$, $V$ be the unitary representation for the CP map
${\cal F}$ and $VU$ be the unitary representation for the CP map ${\cal G}$.
Let $|\Psi\ra_{ss'a}$ be the combined pure state of the system $(s+s')$ 
(after purification) and ancilla. Now in the enlarged Hilbert space we have
the sequence of unitary transformations and direct unitary transformation given
by 
\begin{eqnarray}
|\Psi\ra_{ss'a} &\rightarrow & |\Psi'\ra_{ss'a}=U|\Psi\ra_{ss'a}
 \rightarrow |\Psi''\ra_{ss'a}=V|\Psi' \ra_{ss'a} \nonumber\\
|\Psi\ra_{ss'a} &\rightarrow & |\Psi''\ra_{ss'a}=VU|\Psi\ra_{ss'a}.
\end{eqnarray}

The relative phase shifts, respectively, are given by
\begin{eqnarray}
\Phi_{12} =  {\rm Arg}[ {_{ss'a}}\la \Psi |\Psi'\ra_{ss'a}] \nonumber\\
\Phi_{23} =  {\rm Arg}[ {_{ss'a}}\la \Psi' |\Psi''\ra_{ss'a}] 
\nonumber\\
\Phi_{13} =  {\rm Arg}[ {_{ss'a}}\la \Psi |\Psi''\ra_{ss'a}].
\end{eqnarray}

Let us calculate the quantity $\Phi_{12}+ \Phi_{23}- \Phi_{13}$  which is
nothing but phase difference between the relative phases acquired in 
sequence of quantum operations and the relative phase in direct quantum 
operation. It is given by
\begin{eqnarray}
 \Phi_{12}+ \Phi_{23}- \Phi_{13} = {\rm Arg} \Delta^{(3)}= 
{\rm Arg} \la \Psi |\Psi'\ra \la \Psi' |\Psi''\ra \la \Psi'' |\Psi \ra.
\end{eqnarray}
The object in the rhs is nothing but the argument of a three-point Bargmann 
invariant $\Delta^{(3)}$ which is a complex number, in general, and it 
is non-zero. Hence, the relative phase shift during a 
sequence of quantum operations is non-additive in nature. 
In terms of Kraus operators and density matrices one can express (20) as
\begin{eqnarray}
 \Phi_{12}+ \Phi_{23}- \Phi_{13} =  
{\rm Arg} 
{\rm Tr}_s(\rho_s E_0) {\rm Tr}_s(\rho_s' F_0){\rm Tr}_s(\rho_s'' G_0).
\end{eqnarray}
This implies that a quantum
system undergoing a sequence of CP maps remembers its history through these 
relative phases. This property
might be explored further in assigning memory to quantum channels. For example,
if some quantum alphabets are send across a channel, then by looking at 
the relative phase shift one can know if these states have undergone desired 
quantum operation or there has been some unwanted operations in between.
\\

\section{In-phase quantum channel}

Our notion of relative phase shift helps us to define what is called a 
``in-phase'' quantum channel, i.e., a channel through which if quantum signals
are sent then there will be no relative phase shift of the output states with 
respect to the input states. This definition is in the spirit of ``in-phase''
condition of Pancharatnam for a pure quantum state where it does not acquire
any phase if the inner-product between the initial and final state is real 
and positive. 

Let $\rho_i$ be a set of alphabets of pure state density
operators and $p_i$ be the probability distributions with the mixed state
$\rho =\sum_i p_i \rho_i$ such that $\{p_i, \rho_i\}$ denote the ensemble
of input states. The quantum channel having a unitary representation is
``in-phase'' channel if there exists an initial state of the ancilla
$|A\ra$ such that the quantity
${\rm Tr}_{s} \big[ \rho_s \la A|U|A\ra \big]$ is real and positive.

In fact one can use the above condition and obtain the parallel transport 
condition for mixed states undergoing a CP map. This definition will tell 
us how well a channel preserves phase of the signals. To illustrate this, 
let us consider the depolarization channel
\cite{preskill} acting on a qubit in the initial state
$\rho=\frac{1}{2}(I+ {\bf r}\cdot \sigma)$, where
${\bf r}=(x,y,z)$ is the Bloch vector with the length $|{\bf r}|\leq 1$,
$\sigma = (\sigma_{x},\sigma_{y},\sigma_{z})$ are the
standard Pauli matrices, and $I$ is the $2\times 2$ unit matrix. We can model
this with the Kraus operators
\begin{eqnarray} 
m_{0} = \sqrt{1-p}~I , & & 
m_{1} = \sqrt{p/3}~\sigma_{x} , \nonumber \\ 
m_{2} = \sqrt{p/3}~\sigma_{y} , & & 
m_{3} = \sqrt{p/3}~\sigma_{z}   
\label{eq:depolkraus}
\end{eqnarray} 
that map $\rho \rightarrow \rho' = \frac{1}{2} (I+{\bf r}'  
\cdot \sigma)$. Here, $m_{1}$, $m_{2}$,  
and $m_{3}$, correspond to bit flip, both bit and phase flip,  
and phase flip, respectively. Here, $p$ is the probability that  
one of these errors occurs and it determines the shrinking factor  
$|{\bf r}'|/|{\bf r}|=(1-4p/3)$ of the Bloch vector. 
If the qubit is exposed to the depolarization channel then we have
\begin{eqnarray}   
{\rm Tr}_s(\rho_s \la A|U|A\ra)= {\rm Tr}_{s}(\rho_s E_{0}) = \sqrt{1-p}. 
\end{eqnarray} 
This quantity is real and positive, hence the depolarizing channel is a
in-phase channel. The relative phase shift is zero. Thus the channel only 
reduces the visibility by the factor $\sqrt{1-p}$. The absence of phase 
shifts can be understood from the fact that the depolarization channel  
only shrinks the length of the Bloch vector.  
 
\section{Examples}

\subsection{Phase shift under conditional unitary operator}

In this section we will describe the relative phase shift during
a quantum operation whose unitary representation is a conditional one
in the extended Hilbert space. Let the initial state of the system is
$\rho_s= \sum_k p_k {\rho_k}_s$ with each $\rho_k = |\psi_k\ra\la \psi_k|$ 
and ancilla state is $\rho_a= |A\ra\la A|$. The conditional unitary operator
may be written as $U= \sum_i P_i \otimes u_i$ with $P_i=|i\ra \la i|$ as
one-dimensional projectors in ${\cal H}_s$ and $u_i$ as unitary operators 
in ${\cal H}_a$. One can calculate the relative phase shift either 
from the Kraus operators which are given by $F_{\mu}=
\sum_i P_i \la \mu|u_i|A\ra$. Alternately, we can calculate it from the
expression (17) as

\begin{eqnarray} 
\Phi & = & {\rm Arg}  
{\rm Tr}_{s} \big[ \rho_s \la A|U|A\ra \big]= 
{\rm Arg} \sum_i {\rm Tr}_s (\rho_s P_i) \la A|u_i|A\ra \nonumber\\
&=& \tan^{-1} \bigg[\frac{\sum_k p_k \sum_i |c_i^{(k)}|^2 {\rm Im}\la A|u_i|A\ra}
{\sum_k p_k \sum_i |c_i^{(k)}|^2 {\rm Re} \la A|u_i|A\ra} \bigg],
\end{eqnarray} 
where we have used $|\psi_k \ra= \sum_i c_i^{(k)} |i\ra$. Thus, 
to have a non-zero phase shift it must hold that at least for some $i=l$,
${\rm Im} \la A|u_l|A\ra \not=0$ and all others may be zero.
If for all $i$, the condition ${\rm Im} \la A|u_l|A\ra =0$ holds 
then this channel will be a ``in-phase'' quantum channel.

\subsection{Phase change during randomization}

Here, we will discuss the relative phase shift when a 
qubit undergoes randomization, i.e., a pure qubit state becomes completely
mixed via a quantum operation ${\cal E}$ as given by

\begin{eqnarray}
|\psi\ra \la \psi| &\rightarrow& {\cal E}(|\psi\ra\la \psi|)
= \frac{I}{2},
\end{eqnarray}
where the Kraus operators are $\{E_{\mu} \}=\{ \frac{I}{2}, \frac{\sigma_x}{2},
\frac{i\sigma_y}{2},\frac{\sigma_z}{2} \}$. A unitary representation of the
above quantum operation is
\begin{equation}
U= I\otimes P_0 + \sigma_x \otimes P_1 + i\sigma_y \otimes P_2 + 
\sigma_z \otimes P_3
\end{equation}
with initial state of the ancilla being $|A\ra= \frac{1}{2}(|0\ra
+|1\ra+|2\ra+|3\ra) \in {\cal H}^4$. Again using the expression (17) 
one can calculate the relative phase shift as
\begin{eqnarray} 
\Phi & = & {\rm Arg}  
{\rm Tr}_{s} \big[ \rho_s \la A|U|A\ra \big] \nonumber\\
&=& \tan^{-1} \bigg[\frac{1+ 2 {\rm Im}(\alpha \beta^*)}
{1+ 2 {\rm Re}(\alpha \beta^*)+ (|\alpha|^2 - |\beta|^2)} \bigg].
\end{eqnarray}

Contrary to the usual believe that during a complete randomization, the phase
of a qubit undergoes random changes and becomes maximally mixed, there is
indeed a definite relative phase change during randomization.
This could be useful, for example, in understanding the relative phase shift 
in quantum teleportation channel where
a pure quantum state become completely randomized after Bell-state measurement
but before sending the classical communication.
This will be explored further in future.

\section{Conclusion}
In this work, starting with general notion of relative phases in quantum 
mechanics, we have presented a generalization of the notion  
of relative phase shift when a quantum system undergoes a 
quantum operation described by completely positive maps. In the 
enlarged Hilbert space (so called `Church of the large Hilbert space')
where everything is pure, the notion of relative phase shift coincides
with the Pancharatnam phase shift. We have shown how this phase shift
during quantum operation depends on the initial state of the ancilla.
Further, We have shown how the relative phase shift during a sequence of
quantum operation is non-additive in nature.
This allows us to introduce the notion of ``in-phase quantum channels''
and we gave one example of such a channel. It is hoped that this ideas will 
be useful in the context of quantum information and communication and 
trigger new experiments on relative phases for quantal systems 
exposed to environmental interactions. Much more work needs to be done
to understand the differential geometric structure of geometric phase 
under CP map and its unitary equivalence.\\

\vskip 0.2 cm
{\bf Acknowledgments:} 
Part of the work has been carried out in collaboration with
M. Ericson, E. Sj{\"o}qvist, J. Br{\"a}nnlund, and D. K. L. Oi.
It is a pleasure to thank my collaborators.
I also thank Y. S. Kim and H. Brandt for inviting me to give a talk in 
the First Feynman Festival on Quantum Computing, held from August 23-28th, 
2002 at University of Maryland, College Park, USA.


\vskip 0.3 cm  

\begin{thebibliography}{99} 

\bibitem{feynman} R. P. Feynman, R. B. Leighton,
and M. Sands. {\it Feynman Lectures on Physics, vol III},
Addison-Wesley, Reading, Mass., 1965.

\bibitem{panch} S. Pancharatnam, 
Proc. Indian Acad. Sci. A {\bf 44}, 247 (1956). 

\bibitem{berry84} M.V. Berry,  
Proc. R. Soc. London Ser. A {\bf 392}, 45 (1984). 

\bibitem{aharonov87} Y. Aharonov and J.S. Anandan,  
Phys. Rev. Lett. {\bf 58}, 1593 (1987). 

\bibitem{samuel88} J. Samuel and R. Bhandari,  
Phys. Rev. Lett. {\bf 60}, 2339 (1988). 

\bibitem{aitchison92} I.J.R. Aitchison and K Wanelik, 
Proc. R. Soc. London Ser. A {\bf 439}, 25 (1992); 

\bibitem{ms}N. Mukunda and R. Simon, 
Ann. Phys. (N.Y.) {\bf 228}, 205 (1993).  

\bibitem{akp} A. K. Pati, 
Phys. Rev. A {\bf 52}, 2576 (1995); 

\bibitem{akp1} A.K. Pati, J. Phys. A {\bf 28}, 2087 (1995).
 
\bibitem{uhlmann86} A. Uhlmann, 
Rep. Math. Phys. {\bf 24}, 229 (1986). 

\bibitem{sjoqvist00} E. Sj\"{o}qvist, A.K. Pati, A. Ekert,  
J.S. Anandan, M. Ericsson, D.K.L. Oi, and V. Vedral,  
Phys. Rev. Lett. {\bf 85}, 2845 (2000). 

\bibitem{eric} M. Ericsson, A. K. Pati, E. Sj{\"o}qvist, J. Br{\"a}nnlund, 
and D. K. L. Oi, Quant. Ph, 0206063 (2002).

\bibitem{slater} P. B. Slater, Quant. Ph., 0112054 (2001).

\bibitem{slater1} P. B. Slater, Lett. Math. Phys. {\bf 60(2)}, 123 (2002).

\bibitem{jones00} J.A. Jones, V. Vedral, A. Ekert and G. Castagnoli,  
Nature {\bf 403}, 869 (1999); 

\bibitem{ekert} A. Ekert, M. Ericsson, P. Hayden,  
H. Inamori, J.A. Jones, D.K.L. Oi, and V. Vedral,  
J. Mod. Opt. {\bf 47}, 2501 (2000);  

\bibitem{falci} Giuseppe Falci, Rosario Fazio, G. Massimo Palma,  
Jens Siewert, Vlatko Vedral, Nature {\bf 407}, 355 (2000); 

\bibitem{nazir} A. Nazir, T. P. Spiller, W. J. Munro, 
Phys. Rev. A {\bf 65}, 042303 (2002).  

\bibitem{gbv} I. Fuentes-Guridi, S. Bose, and V. Vedral,
Phys. Rev. Lett. {\bf 85}, 5018 (2000).

\bibitem{barry} B. C. Sanders, H. de Guise, S. D. Bartlett, W. Zhang,
Phys. Rev. Lett. {\bf 86}, 369 (2001).
 
\bibitem{rb} R. Bhandari, Phys. Rev. Lett. {\bf 89}, 268901 (2002). 

\bibitem{anandan} J. S. Anandan,  E. Sj\"{o}qvist, A.K. Pati, A. Ekert, 
M. Ericsson, D.K.L. Oi, and V. Vedral, 
Phys. Rev. Lett. {\bf 89}, 268902 (2002). 

\bibitem{marie} M. Ericsson, E. Sj{\"o}qvist, J. Br{\"a}nnlund, D. K. L. Oi,
and A. K. Pati, Quant. Ph, 0205160 (2002).

\bibitem{fpn} J. G. P. Faria, A. F. R. T. Piza, and
M. C. Nemes, Quant. Ph., 0205146 (2002).

\bibitem{nc} M. Nielsen and I. Chuang, {\it Quantum Computation and
Quantum Information}, Cambridge University Press, 2000.

\bibitem{kraus83} K. Kraus, {\it States, Effects and Operations} 
(Springer-Verlag, Berlin, 1983). 

\bibitem{preskill} J. Preskill, Lecture notes,    
{\it www.theory.caltech.edu/people/preskill/ph229.} 




\end{thebibliography}
\end{document}